\documentclass[useAMS,usenatbib,twocolumn]{mn2e}
\usepackage{amsmath}
\usepackage{graphicx}
\usepackage{natbib}
\usepackage{color}
\usepackage{hyperref}

 \providecommand{\adsurl}[1]{\href{#1}{ADS}}

\newcommand{\ltsima} {$\; \buildrel < \over \sim \;$}
\newcommand{\gtsima} {$\; \buildrel > \over \sim \;$}
\newcommand{\lta} {\lower.5ex\hbox{\ltsima}}
\newcommand{\gta} {\lower.5ex\hbox{\gtsima}}

\newcommand{\jcap}{JCAP}

\newcommand{\invh}{{\sf{h^{-1}}}}

\title[The length of the low-redshift standard ruler]{The length of the low-redshift standard ruler}
\author[Licia Verde, Jos\' e Luis Bernal, Alan Heavens, Raul Jimenez]{Licia Verde$^{1,2,3,4,5}$, Jos\' e Luis Bernal$^{1,6}$, Alan Heavens$^7$
, Raul Jimenez$^{1,2,3,4}$ \\
$^1$ICC, Instituto de Ciencias del Cosmos, Universitat de Barcelona, IEEC-UB, Mart{\'\i} i Franqu{\`e}s 1, E-08028, Barcelona, Spain\\
$^2$ICREA, Pg. Llu's Companys 23, 08010 Barcelona, Spain\\
$^3$Radcliffe Institute for Advanced Study, Harvard University, MA 02138, USA\&\\ 
$^4$Institute for Theory and Computation, Harvard-Smithsonian Center for Astrophysics, 60 Garden Street, Cambridge, MA 02138, USA\\
$^5$Institute of Theoretical Astrophysics, University of Oslo, 0315 Oslo, Norway\\
$^6$Dept. de  F\' isica Qu\` antica i Astrof\' isica, Universitat de Barcelona, Mart\' i i Franqu\` es 1, E08028 Barcelona, Spain\\
$^7$Imperial Centre for Inference and Cosmology, Imperial College, Blackett Laboratory, Prince Consort Road, London SW7 2AZ, U.K.}

\date{Accepted ;  Received ; in original form }

\begin{document}
\maketitle

\begin{abstract}
Assuming the existence of  standard rulers, standard candles and standard clocks, requiring only the cosmological principle, a metric theory of gravity, a smooth expansion history, and using  state-of-the-art observations, we determine the length of the ``low-redshift standard ruler". The data we use are  a compilation of recent Baryon acoustic oscillation data (relying on the standard ruler), Type 1A supernov\ae\ (as standard candles), ages of early type galaxies  (as standard clocks) and local determinations of the Hubble constant (as a local anchor of the cosmic distance scale).  In a standard   $\Lambda$CDM cosmology  the ``low-redshift standard ruler" coincides with the sound horizon at radiation drag, which can also be determined --in a model dependent way-- from CMB observations. However, in general, the two quantities need not coincide. We obtain  constraints on the length of the  low-redshift standard ruler: 
 $r^h_{\rm s}=101.0 \pm 2.3 h^{-1}$ Mpc, when using only Type 1A supernov\ae\ and Baryon acoustic oscillations, and  $r_{\rm s}=150.0\pm 4.7 $ Mpc when using clocks to set the Hubble normalisation, while  $r_{\rm s}=141.0\pm 5.5 $ Mpc when using the local Hubble constant determination (using both yields $r_{\rm s}=143.9\pm 3.1 $ Mpc).  
 
The low-redshift determination of the standard ruler has an error which is competitive with  the model-dependent determination from cosmic microwave background measurements made with the {\em Planck} satellite, which assumes it is the sound horizon at the end of baryon drag.
\end{abstract}


\section{Introduction}

We build on the idea presented in \cite{Sutherland2012} and \cite{Heavens:2014rja}  that relatively low redshift measurements of the cosmic expansion history $H(z)$, can be used, in combination with measurements of the  Baryon Acoustic Oscillation (BAO)  feature, to determine the length of a standard ruler in a model-independent way. Supernovae type 1A are standard(izable) candles yielding a luminosity-distance -- redshift relation. The BAO  feature is probably the best-understood standard ruler in the Universe. However, it has the drawback that  the comoving length of the ruler,  the sound horizon at radiation drag  $r_{\rm s}$,  is  usually calibrated at $z > 1000$  relying on  Cosmic Microwave Background  (CMB) observations and  theoretical assumptions.  Without knowing the length of the ruler  or the brightness of the candles or the Hubble parameter, these probes can only give relative measurements of the expansion history. The quantities  $r_{\rm s}$ and $H_0$ provide absolute scales for distance measurements (anchors) at opposite ends of the observable Universe. But while the CMB $r_{\rm s}$ determination depends on several assumptions (standard gravity, standard radiation content, negligible isocurvature perturbations,  standard scaling of matter and radiation components, negligible  early dark energy  etc.), local determinations of the expansion rate are cosmology-independent.  Alternatively  standard clocks \citep{Jimenez_C} can be used, representing objects whose age is determined by established physics, and whose formation time is sufficiently early that scatter amongst formation times is negligible in the present cosmological context. Standard clocks provide (absolute) measurements of $H(z)$.

Even relative measurements of the expansion history, from observations of Type 1A supernov\ae, in combination with measurements of the BAO feature can yield a  constraint  on the low-redshift standard ruler,
$r^h_{\rm s}$,  which is  the ruler length in units of $h^{-1}$Mpc.  An {\em absolute} distance scale can be provided by adding a constraint on $h$ such as that provided by $H_0$ or clocks, in which case observations of the BAO feature can be used to determine the  absolute length of the {\it low-redshift standard ruler},  $r_{\rm s}$, in units of Mpc. The  importance of this  scale is that it is a key theoretical prediction of cosmological models, depending on the sound speed and expansion rate of the Universe at early times, before matter and radiation decouple. However the low-redshift standard ruler is a direct measurement, which will survive even if the standard cosmological model and standard assumptions about early-time physics  do not. Since the analysis of \cite{Heavens:2014rja}, new BAO, $H_0$, and cosmic clock data have become available, with improved  statistics, which we consider here.
 
\section{Data and Methodology}

The latest $H_0$ determination is provided by the SH0ES program, reaching a 2.4\% precision, $H_0^{SH0ES}=73.24\pm 1.74$ km s$^{-1}$Mpc$^{-1}$ \citep{RiessH0_2016}. A Gaussian likelihood is assumed. 

The  supernov\ae\  type 1A data are the compilation of  \cite{Betoule14_jla},  binned into 31 redshift intervals between $0$ and $1.3$, equally-spaced in $\log(1+z)$ to yield the distance modulus as function of redshift.  The covariance matrix is supplied for the binned data.  The binning, in conjunction with the central limit theorem, motivates the use of a gaussian likelihood. 
The data is given as measurements of the distance modulus
\begin{equation}
  \mu(z)\equiv m-M =  25+5\log_{10} D_L(z)
  \label{eq:mu}
\end{equation}
where $m$ is the apparent magnitude, $M$ a fiducial  absolute magnitude $M\simeq -19.3$ and $D_L$ the luminosity distance.

Constraints on BAO  are from the following galaxy surveys: Six Degree Field Galaxy Survey (6dF) \citep{Beutler11}, 
the LOWZ and CMASS galaxy samples of the Baryon Oscillation Spectroscopic Survey (BOSS-LOWZ and BOSS-CMASS, respectively, \cite{Cuesta16_bao}, we use the isotropic measurement), and the reanalysed measurements of WiggleZ \citep{WZ} by \cite{Kazin14_wz}. While we take into account  the  correlation among the  WiggleZ measurements we neglect the correlation between  WiggleZ and CMASS. This is motivated by the fact that the WiggleZ-CMASS overlap includes a small fraction of the BOSS-CMASS sample and the correlation is very small, always below 4\% \citep{Beutler16_overlap, Cuesta16_bao}.
BAO data provide measurements of the dilation scale  normalized by the standard ruler length,  $D_V/r_{\rm s}$, where
\begin{equation}
D_V(z) \equiv \left[(1+z)^2D^2_A(z) \frac{cz}{H(z)}\right]^{1/3}.
\end{equation}
If $r_{\rm s}$ is interpreted as  the sound horizon at radiation drag, $r_d(z_d)$, then
\begin{equation}
r_d (z_d)=\int_{z_d}^{\infty}\frac{c_s(z)}{H(z)}dz,
\end{equation}
where $c_s(z)$ is the sound speed.

For the standard clocks, we use galaxy ages determined from analysis of stellar populations of old elliptical galaxies. We assume that the formation time was at sufficiently high redshift that variations in formation time of stars within each galaxy and among galaxies are negligible.  Differential ages, $\Delta t$, then provide estimates of the inverse Hubble parameter as $1/H(z)= dt/dz(1+z)$ and $dt/dz\simeq \Delta t/\Delta z$ for suitable redshift intervals  $\Delta z$. We use the measurements of $H(z)$ obtained by \cite{Moresco16}, who  extend  the previously available compilation to include  both a fine sampling at $0.38<z<0.48$ exploiting the  unprecedented statistics provided by the BOSS Data Release 9, and the redshift range up to $z\sim2$.

As in \cite{Heavens:2014rja}, we parametrise the  expansion history by an inverse Hubble parameter, $\invh(z) \equiv 100 {\rm km}\,{\rm s}^{-1}{\rm Mpc}^{-1}/H(z)$, which is specified at $N=7$ values (nodes) equally-spaced between $z=0$ and $z=1.97$; we linearly-interpolate $\invh(z)$ in  between.  Since the   maximum redshift probed by  supernov\ae\  data is smaller than that probed by clocks, when clocks are not included $N=5$ and the maximum redshift value considered is $z=1.3$. 
This implicitly assumes a smooth expansion history.

Assuming the cosmological principle of homogeneity and isotropy (and thus a FRW metric), the curvature of the Universe ($k=\{1, 0,-1\}$) and $H(z)$ completely specify the metric and the geometric observables  considered here: luminosity distance $D_L$,   and the dilation scale $D_V$ through the angular diameter distance $D_A$. 
The curvature radius of the Universe is  $k \,R_0$ (for $k=\pm1$) and infinity for $k=0$, where  $R_0$ denotes  the present value of the scale factor, and the curvature is $\kappa=c/(R_0 H_0)$. 
If we wish further to assume General Relativity (GR), the  curvature density parameter is given by   $\Omega_k=k [c/(R_0 H_0)]^2=k\kappa^2$ with $c$ the speed of light.
\footnote{In fact recall that
\begin{equation}
r(z) = \frac{c}{R_0 H_0} \int_0^z \frac{dz'}{E(z')} \equiv \frac{c}{R_0 H_0} \tilde r(z),
\end{equation}
where $E(z)\equiv H(z)/H_0$ and $H(z) = a^{-1}da/dt$.  
\begin{equation}
D_A(z) =  (1+z)^{-1}\frac{c}{H_0 \kappa}S_k\left(\kappa \tilde r\right)\,,
\end{equation}
where   $S_k(r)=\sin r,r,\sinh r$ for $k=1,0,-1$ respectively.
For any metric theory of gravity, the angular diameter distance and luminosity distance are related by $D_L=(1+z)^2 D_A$.}

As it is customary   for supernov\ae, we allow an absolute magnitude offset $\Delta M$:  we are assuming the existence of a standard candle, but not its luminosity. 
  Similarly, for the BAO measurements, we assume there is a standard ruler,  which is normally interpreted as the sound horizon at radiation drag, but for the purposes here, it is simply a ruler. 
   
  The parameters are therefore ($r_{\rm s}^h$, $\Omega_k$, $\Delta M$, $\invh(0)$, $\invh(z_1)$,$\ldots,\invh(z_N)$).  Uniform priors are assumed for all parameters. The parameter space is explored via standard Monte Carlo Markov Chain (MCMC) methods.
  
  In Sec. \ref{sec:robustness} we compare this parametrization with a prior on $H(z)$ in five knots, $r_{\rm s}$  and a spline  interpolation.  We also compare results for different sampling techniques: Metropolis Hastings \citep{MH70} and  Affine Invariant sampler \citep{Goodman_mcmc}.

\begin{table*}
\centering
\begin{tabular}{lccccc}
\hline
Data & $r^h_s [h^{-1}\rm Mpc]$ & $r_{\rm s} [\rm Mpc]$&$H_0$  &$\Delta M$ & $\Omega_k = k (c/H_0 R_0)^2$  \\
\hline
SBH & $ 102.0 \pm 2.5 (^{+2.2}_{-2.8}) $ & $ 140.8 \pm 4.9$ &$ 72.8\pm1.8  $ &$0.079 \pm 0.083 $ & $-0.49 \pm 0.64 \,\left(-0.99^{+0.86}_{-0.26}\right) $\\
BH & $  107.2 \pm 7.2 $ & $ 147 \pm 10$ &$73.0 \pm 1.8 $ &N/A & unconstrained\\
SB& $ 101.0\pm 2.3$ & unconstrained &  unconstrained &  unconstrained & $0.07 \pm 0.61 $  \\

CB & $ 103.9 \pm 5.6$ & $ 149.5\pm4.3 $ &$ 69.6\pm 4.2 $ &N/A & unconstrained \\
CSB&$  100.5\pm 1.9$ & $150.0 \pm  4.7$ &$ 67.0\pm 2.5 $ &$ -0.090\pm 0.079 $ & $ 0.36\pm  0.41$\\
CBH&$ 107.2 \pm 3.4 $ & $148.0 \pm 3.9$ &$72.5 \pm 1.7$ &N/A & unconstrained\\
CSBH & $ 102.3 \pm 1.8 $ & $143.9 \pm 3.1$ &$ 71.1\pm 1.5 $ &$ 0.028\pm0.047  $ & $-0.03 \pm 0.31 \left(-0.08^{+0.32}_{-0.28}\right)  $ \\
\hline
SBH & $ 100.7 \pm 1.8 $ & $138.5 \pm  4.3 $ &$ 72.8\pm 1.8 $ &$  0.083\pm 0.061 $ &  flat \\
BH & $ 107.1 \pm 7.2 $ & $ 147\pm 10$ &$73.0 \pm 1.8 $ & N/A  &  flat  \\
SB& $ 101.2\pm 1.8$ & unconstrained &  unconstrained &  unconstrained & flat   \\
CB & $ 103.7 \pm 5.5$ & $ 149.8\pm 4.2 $ &$69.2 \pm 4.0 $ &N/A &  flat  \\
CSB&$  101.4\pm 1.7 $ & $ 148.3 \pm 4.3 $ &$68.5 \pm 2.1 $ &$-0.047 \pm 0.064 $ &  flat \\
CBH&$ 107.4 \pm 3.4 $ & $ 148.0\pm 3.6$ &$ 72.6\pm1.7  $ &N/A &  flat \\
CSBH & $ 102.3 \pm 1.6 $ & $143.9 \pm 3.1$ &$ 71.1\pm 1.4 $ &$0.026 \pm 0.043 $&  flat \\
\hline
\end{tabular}
\caption{Posterior mean and standard deviation for the model parameters.   The curvature radius of the Universe $R_0$ is constrained, independently of General Relativity, but we report it in terms of the GR-specific curvature density parameter $\Omega_k$. The curvature  distribution in some cases is highly non-gaussian: therefore we also report  in parenthesis the maximum of the posterior and the 68\%  highest posterior density interval. When SNe are not included $\Delta M$ is not a parameter (hence the ``N/A'' table entry).  }
\label{Tab:results1}
\end{table*}

\begin{figure*}
\includegraphics[width=0.48 \textwidth]{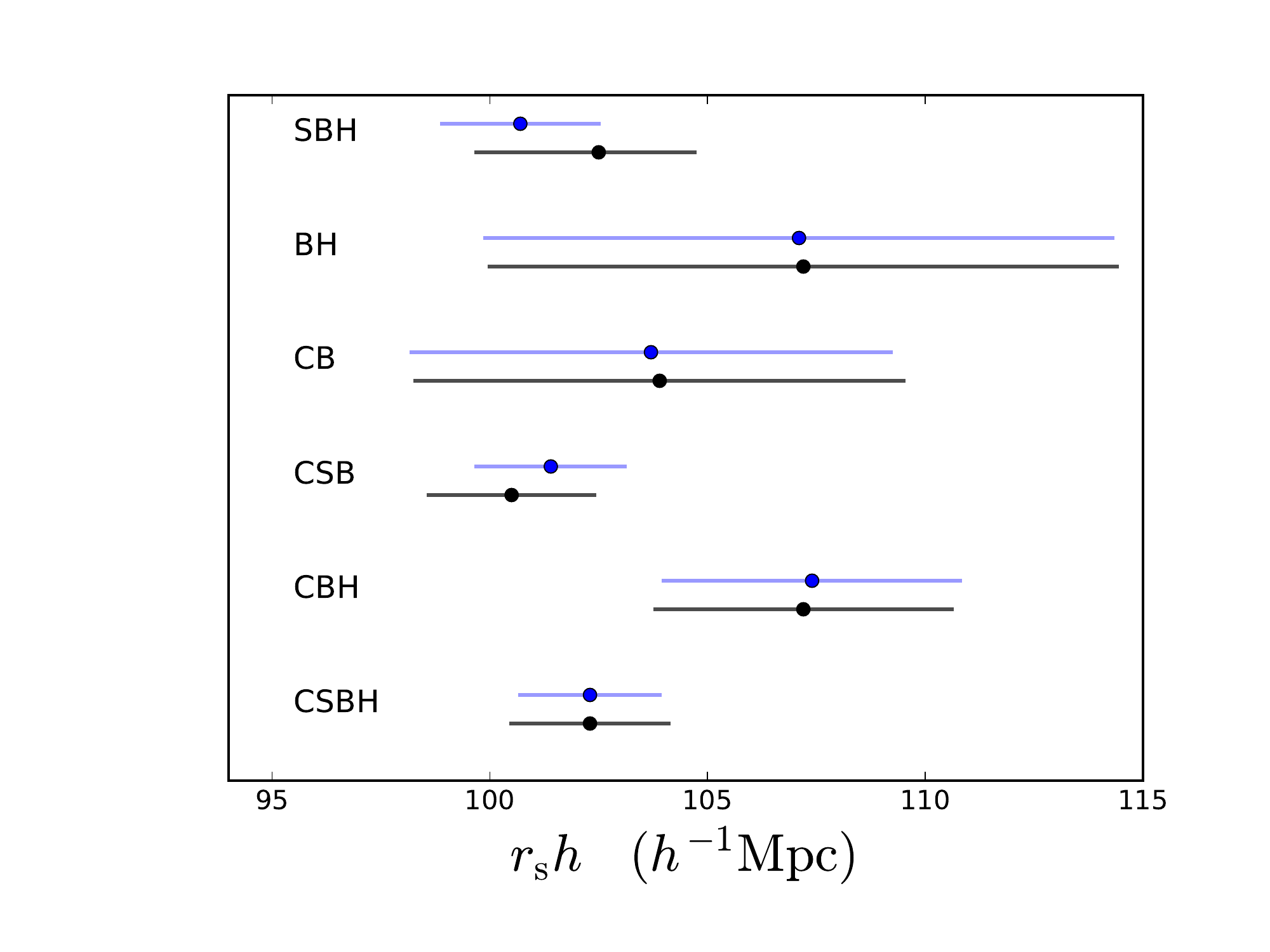}
\includegraphics[width=0.48 \textwidth]{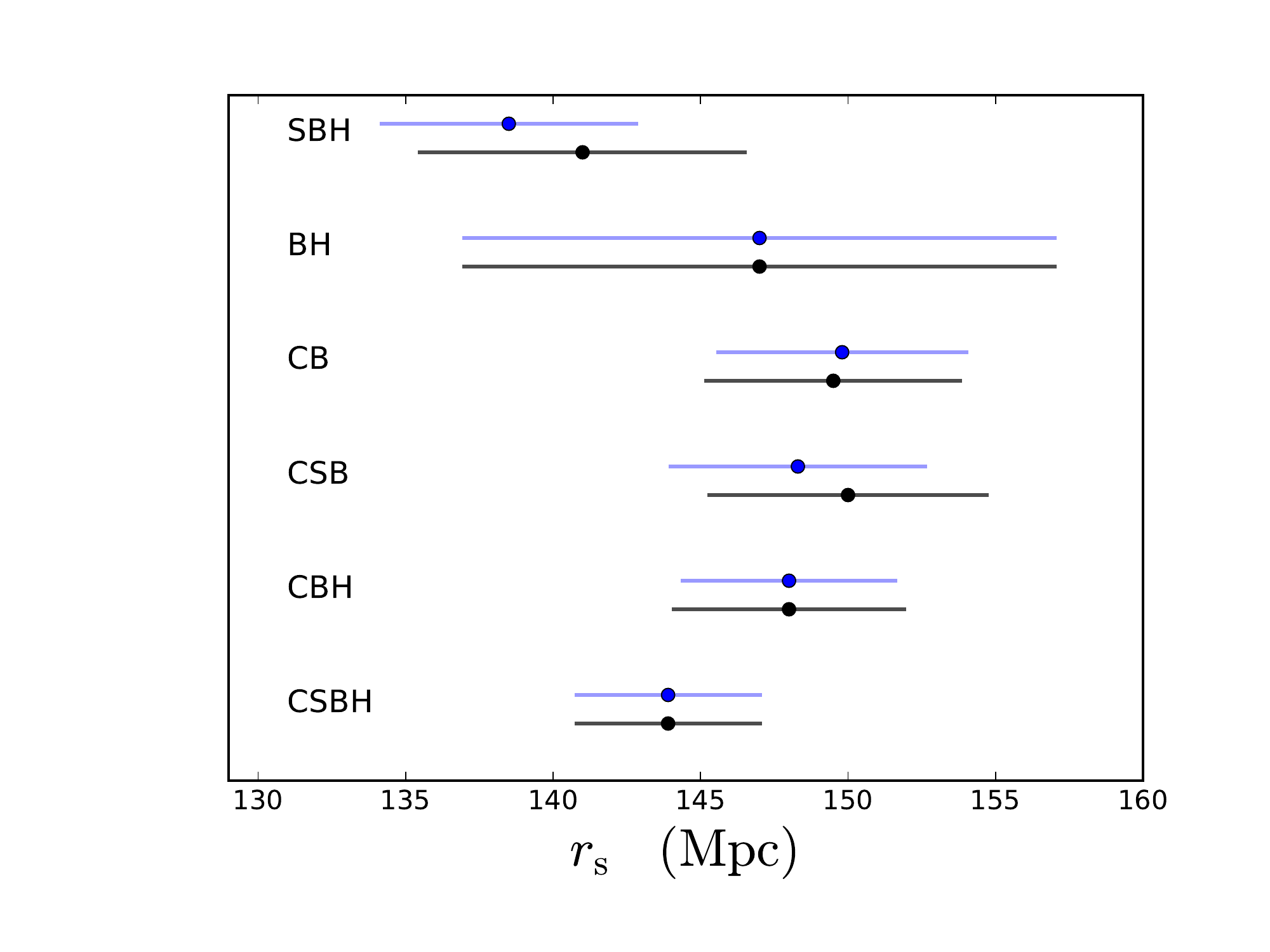}
\caption{At glance: comparison of central values and 1$\sigma$ errors on the $r_{\rm s}^h$ (left) and $r_{\rm s}$ (right) measurements for flat geometry (blue) and marginalizing over the curvature (black). Note the change of the scale in the x-axis in each figure. }
\label{fig:bars1}
\end{figure*}

\section{Results}
In table \ref{Tab:results1} we report the  mean and 68\%  credible regions for the recovered quantities for various combinations of the data:  CSBH indicating Clocks,  supernov\ae, BAOs, local $H_0$ respectively.

The posterior distribution of the curvature parameter is highly non-gaussian, except when both clocks and Supernovae  data are considered or in the SB case;   the curvature  is poorly constrained otherwise, hence in these cases we also report the  maximum of the posterior and the 68\%  highest posterior density interval. 

\begin{figure*}
\includegraphics[width=0.48\textwidth]{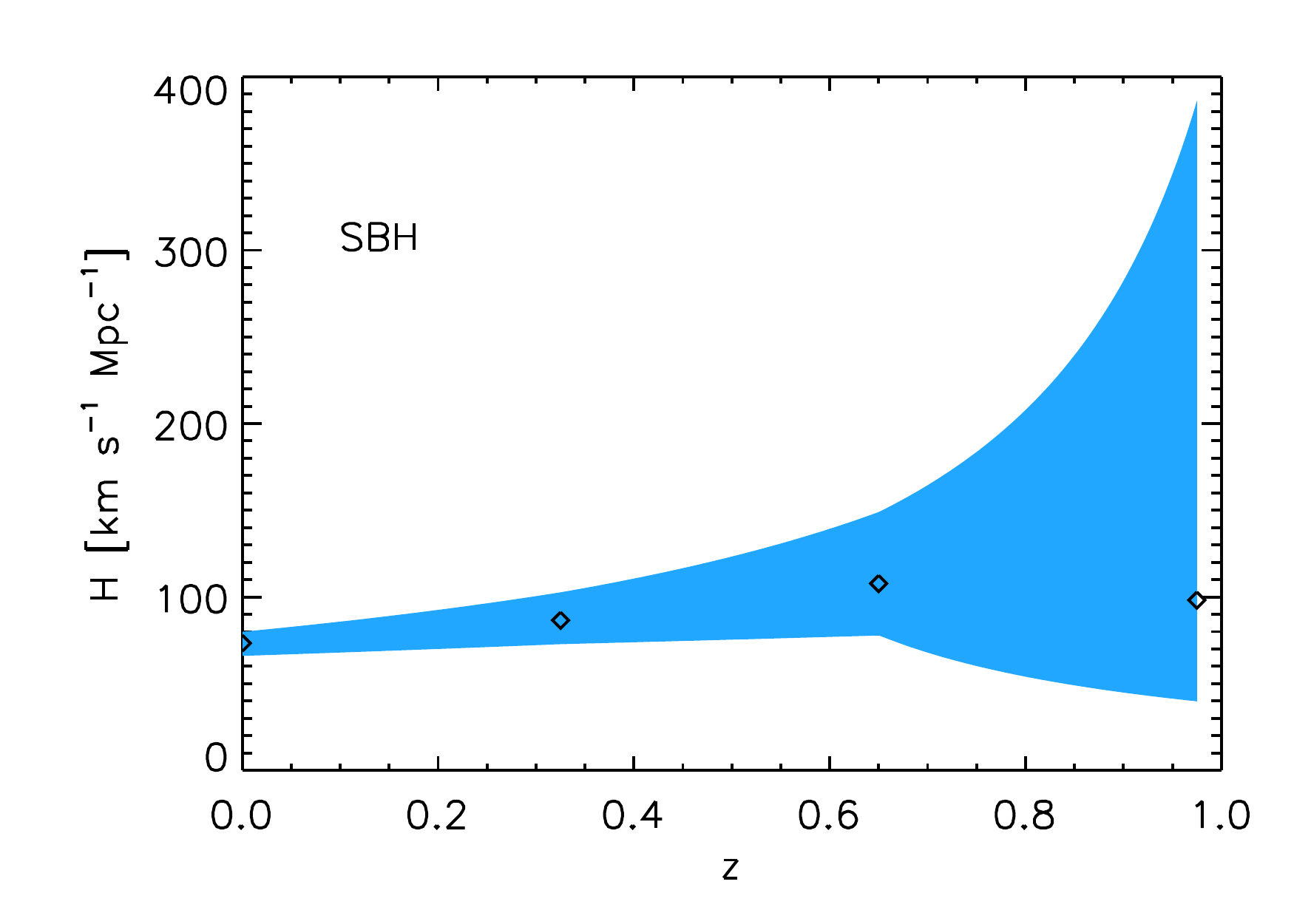}
\includegraphics[width=0.48 \textwidth]{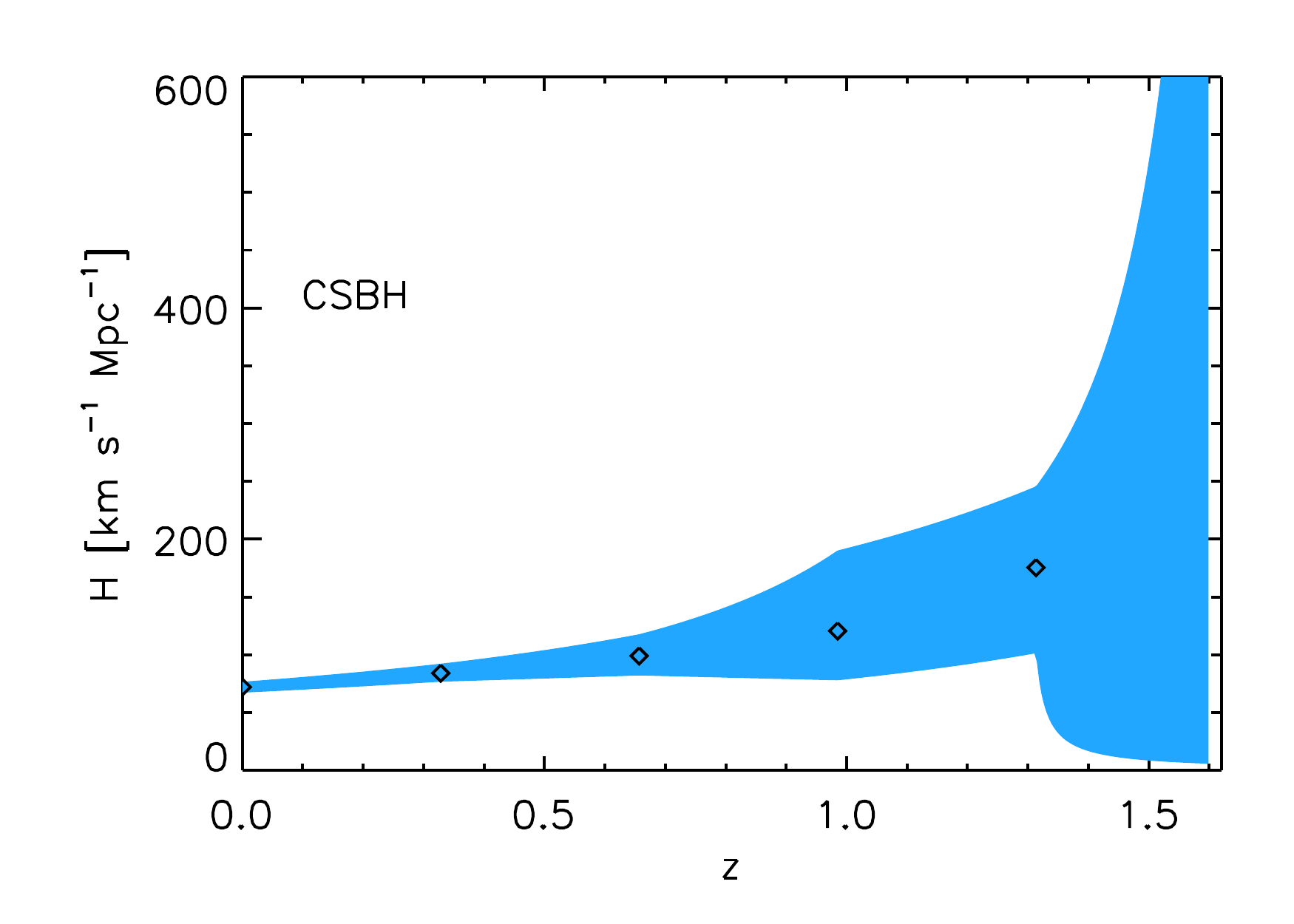}
\caption{Reconstructed expansion history  $H(z)$  (95\% confidence envelop) for two representative dataset combinations: SBH (left) CSBH (right). The last  redshift nodes (one on the left, two on the right) are not shown as there $H(z)$ is poorly constrained. The jagged shape of the envelop is due to the linear interpolation  being performed in $1/H$ while the quantity plotted is $H(z)$.  Symbols represent the best fit values for the reconstruction. }
\label{fig:Hz}
\end{figure*}

The results of Tab.~\ref{Tab:results1} indicate  the following.
\begin{itemize}
\item[-] There is only a mild dependence of the low-redshift standard ruler determination  on curvature. Imposing flatness reduces slightly the error bars, and has no effect when all datasets are considered.  Only in the case of SBH does imposing flatness induce a change of $\sim 1\sigma$ in the low-redshift standard ruler towards lower values.
\item[-] The recovered $H_0$  estimates  cluster around two values: $h\equiv {\sf h}(z=0) \sim 0.73$ obtained when the local  $H_0^{SH0ES}$ is used (as expected);  and  $h\sim 0.68$ when clocks are used, and $0.71$ when both are used.
\item[-] The $H_0$ value obtained by the CSB combination has an error bar of 3.7\%, to be compared with a 2.4\% error for $H_0^{SH0ES}$ and a 3.8\% error for H0LiCOW \citep{H0_holicow}. These two measurements are in agreement at the $2 \sigma$ level  with the CSB value.
\item[-] supernov\ae\  and cosmic clocks data are needed to constrain the curvature. The curvature distribution is highly non-gaussian, unless these data sets are considered.  
\item[-]   without $H_0^{SH0ES}$,   $r_{\rm s}$ tends to be $\sim 149$Mpc,   as expected, $H_0^{SH0ES}$ pulls the recovered $r_{\rm s}$ downwards. 
\item[-]  depending on how extensive the dataset considered is, the error on $r_{\rm s}^h$ varies between 7\% (for BH) to  1.8\% (CSBH), the error on  $r_{\rm s}$ varies between 7\% (for BH) to  2.1\% (CSBH).
\item[-]  while $r_{\rm s}^h$  is better determined  than $r_{\rm s}$ the  recovered value across different data sets is more consistent for   $r_{\rm s}$.
\item[-] $r_{\rm s}^h$ is determined at the $2$\% level with only BAO and Supernovae. In this case the curvature distribution is remarkably more symmetric than for the SBH case.
\end{itemize}

Figure \ref{fig:bars1} offers visual comparisons  of the $r_{\rm s}^h$ and $r_{\rm s}$
measurements, for the flat case and marginalizing over curvature. The CSB combination yields a $r_{\rm s}$ value fully consistent with the Planck mission CMB inferred one,  while the  SBH  determination  yields lower values, which are  still consistent  in the case of the non-flat case but become a  $\sim 2\sigma$ tension (with respect to the Planck value for the $\Lambda$CDM model) when flatness is imposed.

\begin{figure*}
\includegraphics[width=0.33 \textwidth, angle=90]{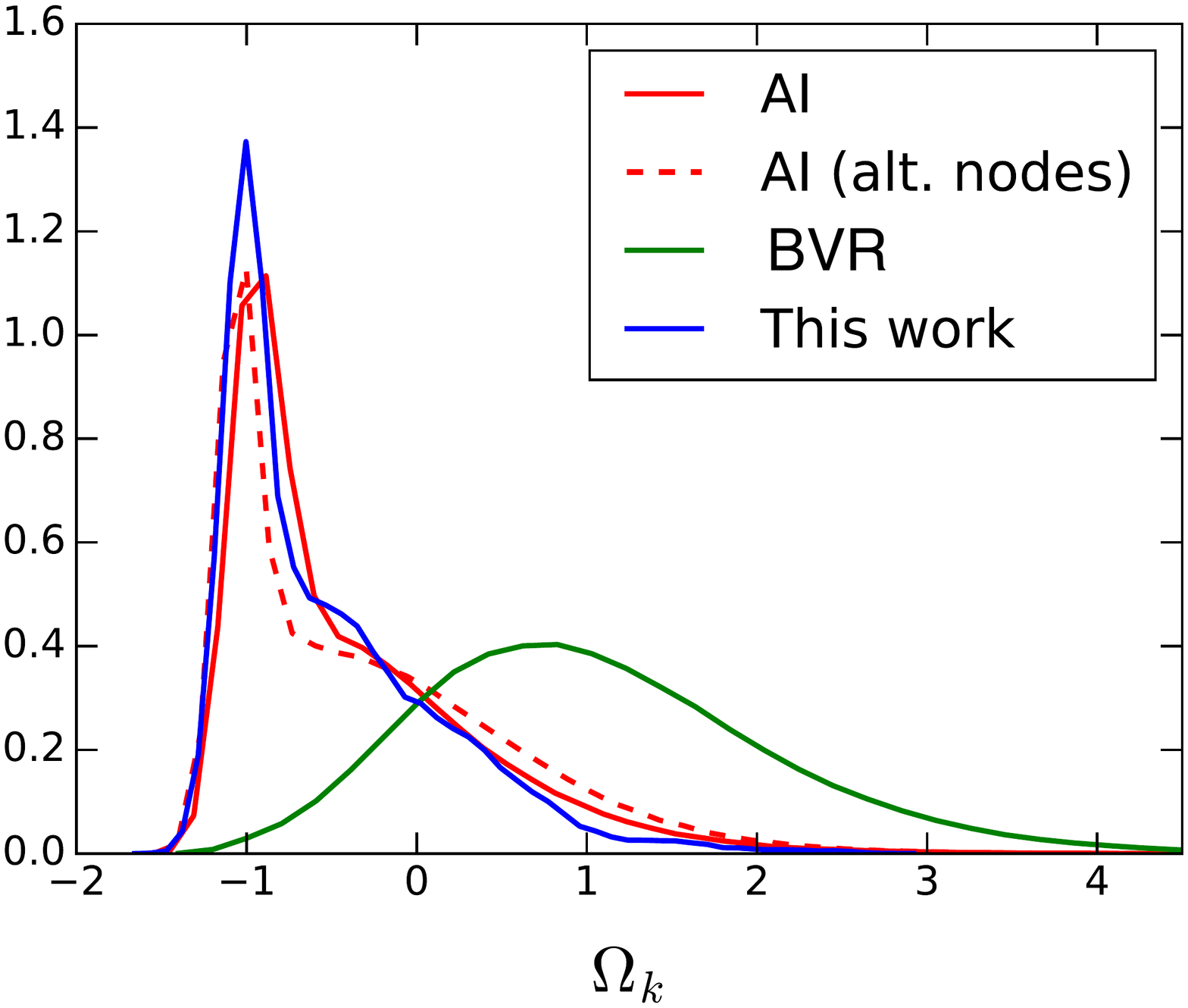}
\includegraphics[width=0.33 \textwidth, angle=90]{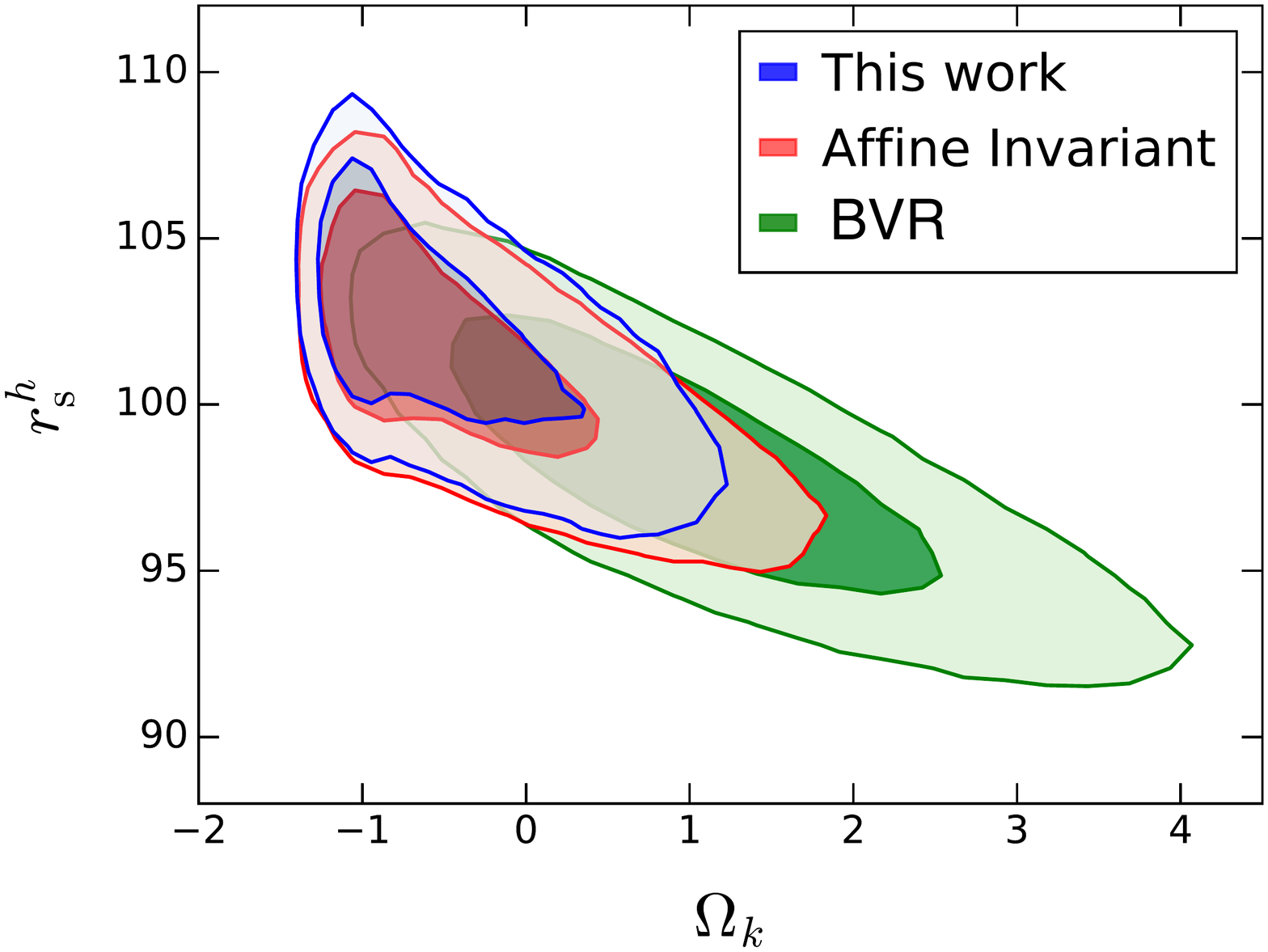}
\caption{Effects of  prior assumptions and MCMC sampling method. We show the comparison of the posterior distributions of $\Omega_k$ (left) and in the $\Omega_k$-$r_{\rm s}^h$ plane (right) obtained  from the same data (SBH)  with different methodologies: this work (blue),  using an Affine Invariant sampler instead of Metropolis Hastings (red) with two choices for the redshift  sampling, the one form this work (solid) and the one from BVR (dashed), and  the approach of  BVR (green), which uses Affine Invariant sampler,  $r_{\rm s}$ and $H(z_i)$ as variables and a spline interpolation of $H(z)$.}
\label{fig:omk_prior}
\end{figure*}

Fig. ~\ref{fig:Hz} shows the  envelope enclosing 95\% of the reconstructed $H(z)$ for two representative data set combinations.  The odd shape of the envelope is due to the fact that  the linear interpolation is being performed in $1/H$ while the quantity plotted is $H(z)$. Symbols represent the best fit $H(z)$ of each redshift.  The highest redshift nodes are poorly constrained and therefore not shown. Also for the CSBH case, the joint distribution of the $\invh(z_n)$  values for the last two redshift nodes.   show a structure indicating a high degree of interdependence between the two  quantities. This does not affect the determination of the standard ruler, as there is no correlation between $r_{\rm s}$ or $r_{\rm s}^h$ and $\invh(z_n)$ for $n \ge 4$.

 \begin{figure*}
\includegraphics[width=0.45\textwidth]{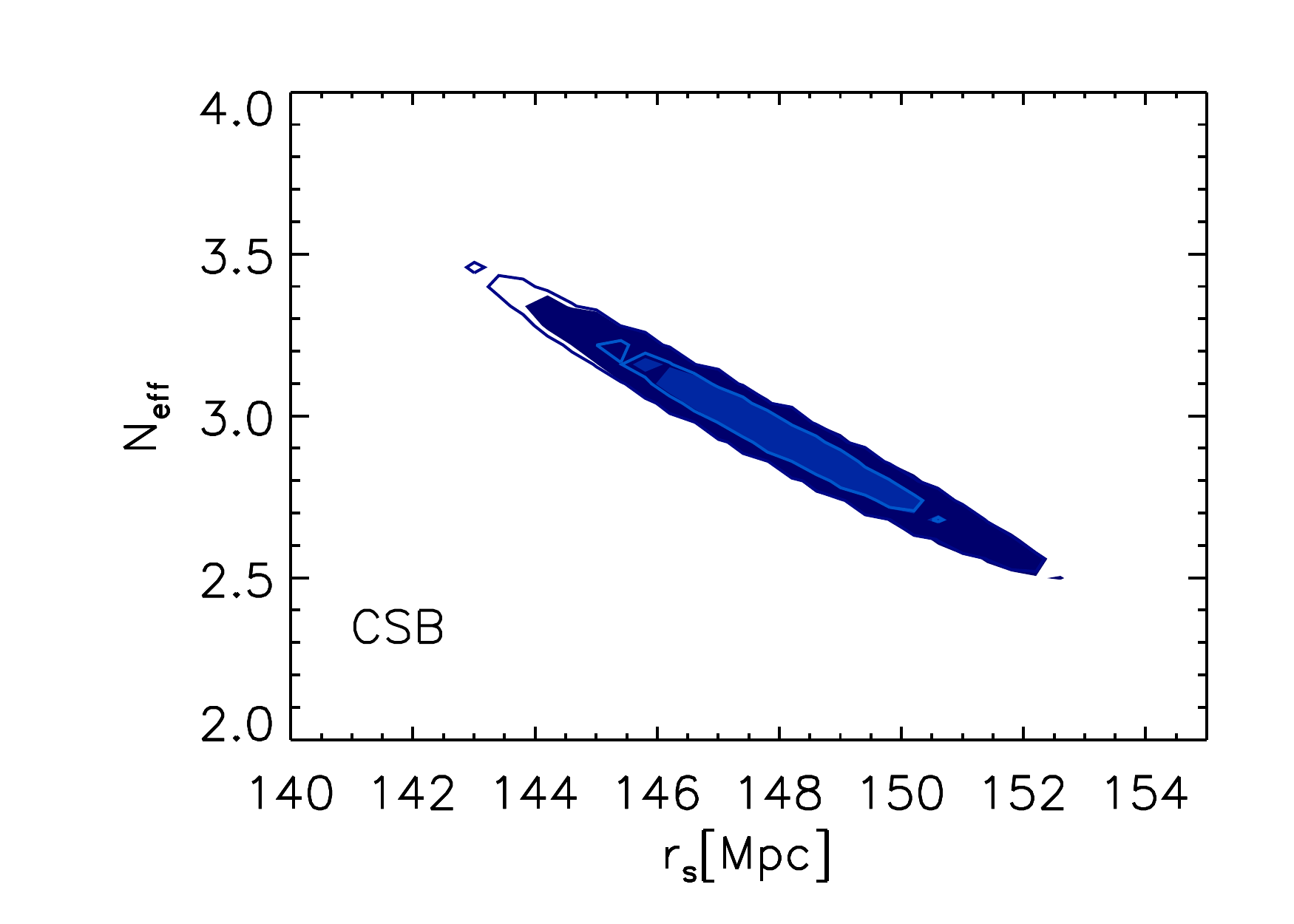}
\includegraphics[width=0.45 \textwidth]{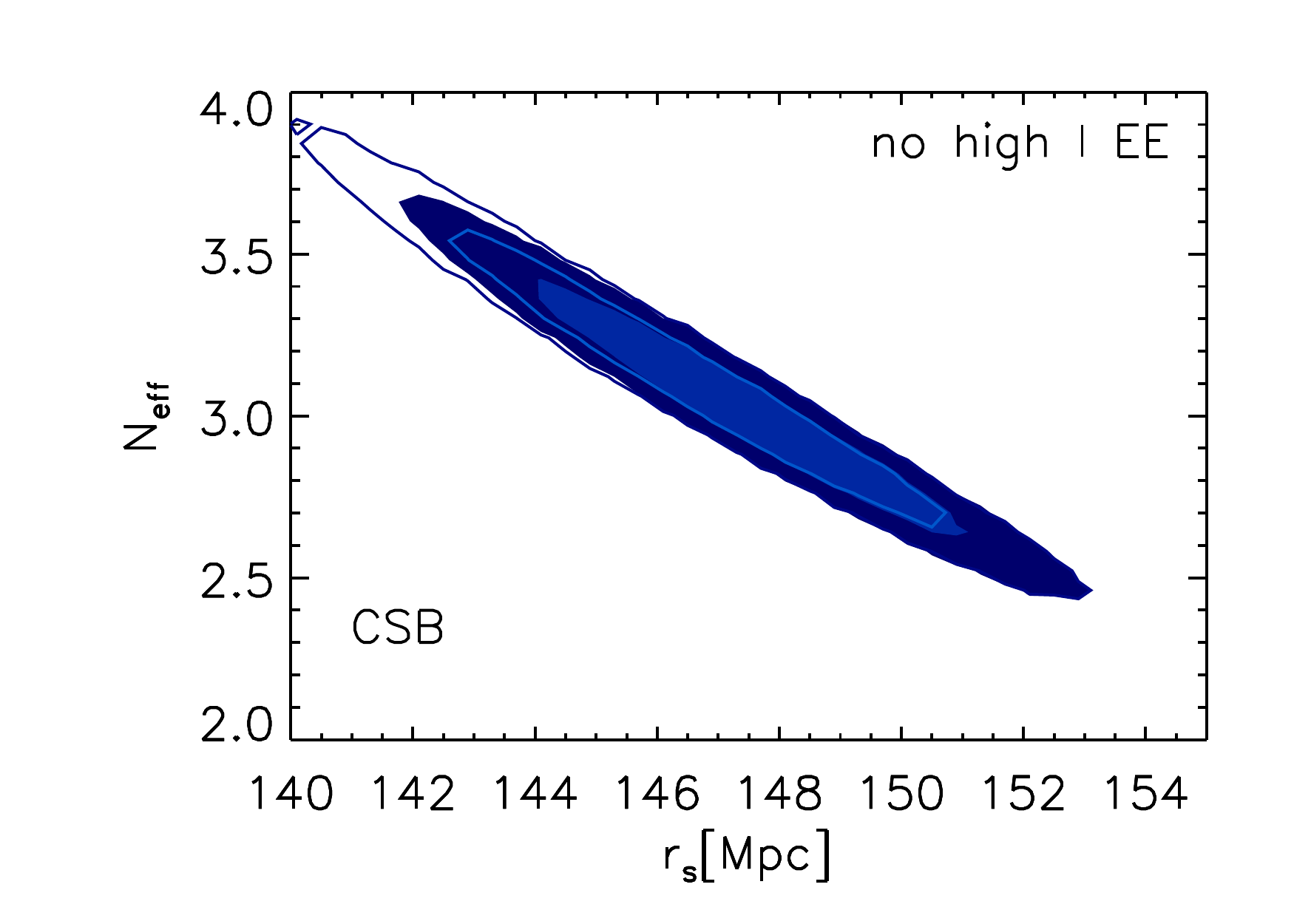}
\includegraphics[width=0.45\textwidth]{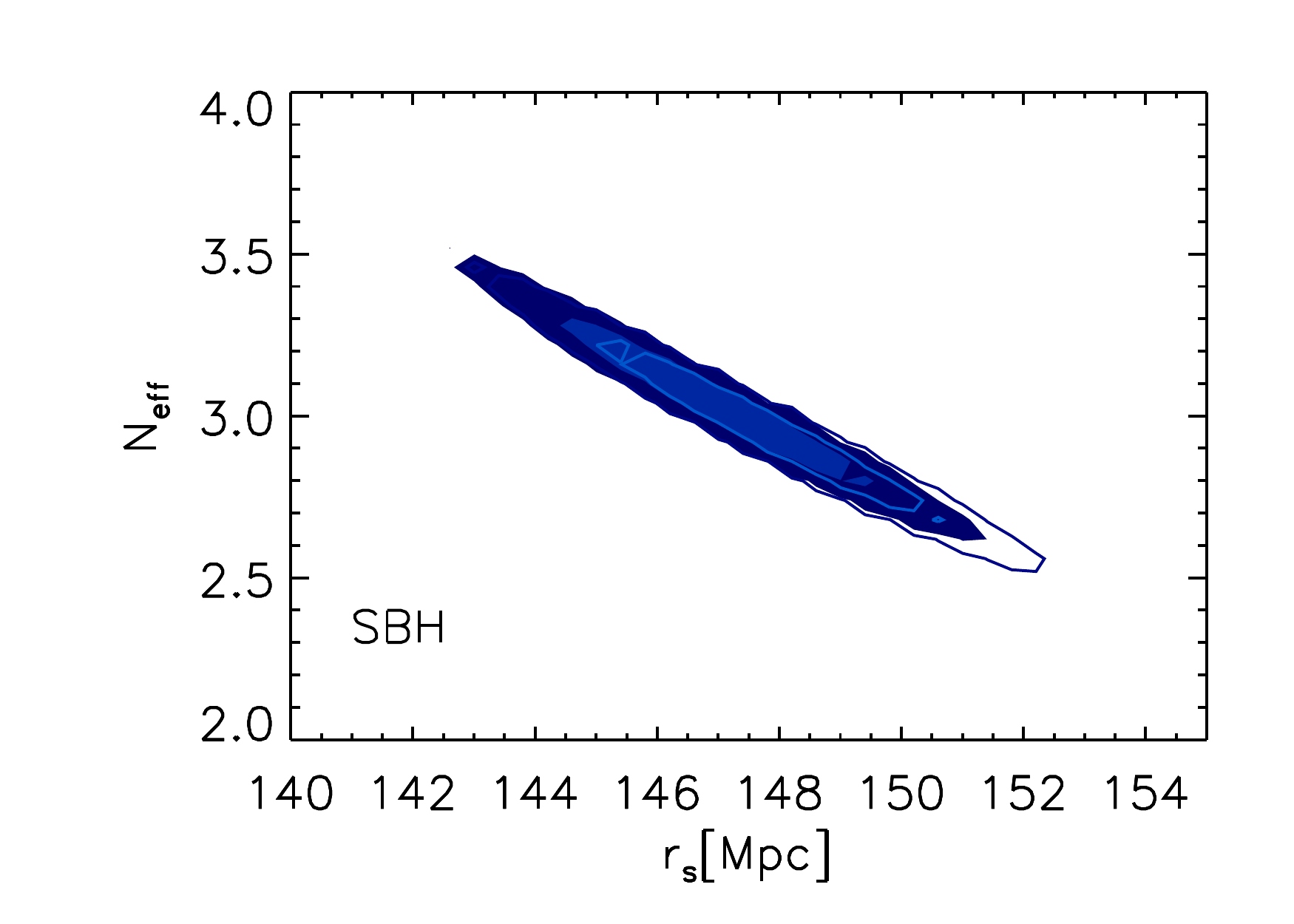}
\includegraphics[width=0.45 \textwidth]{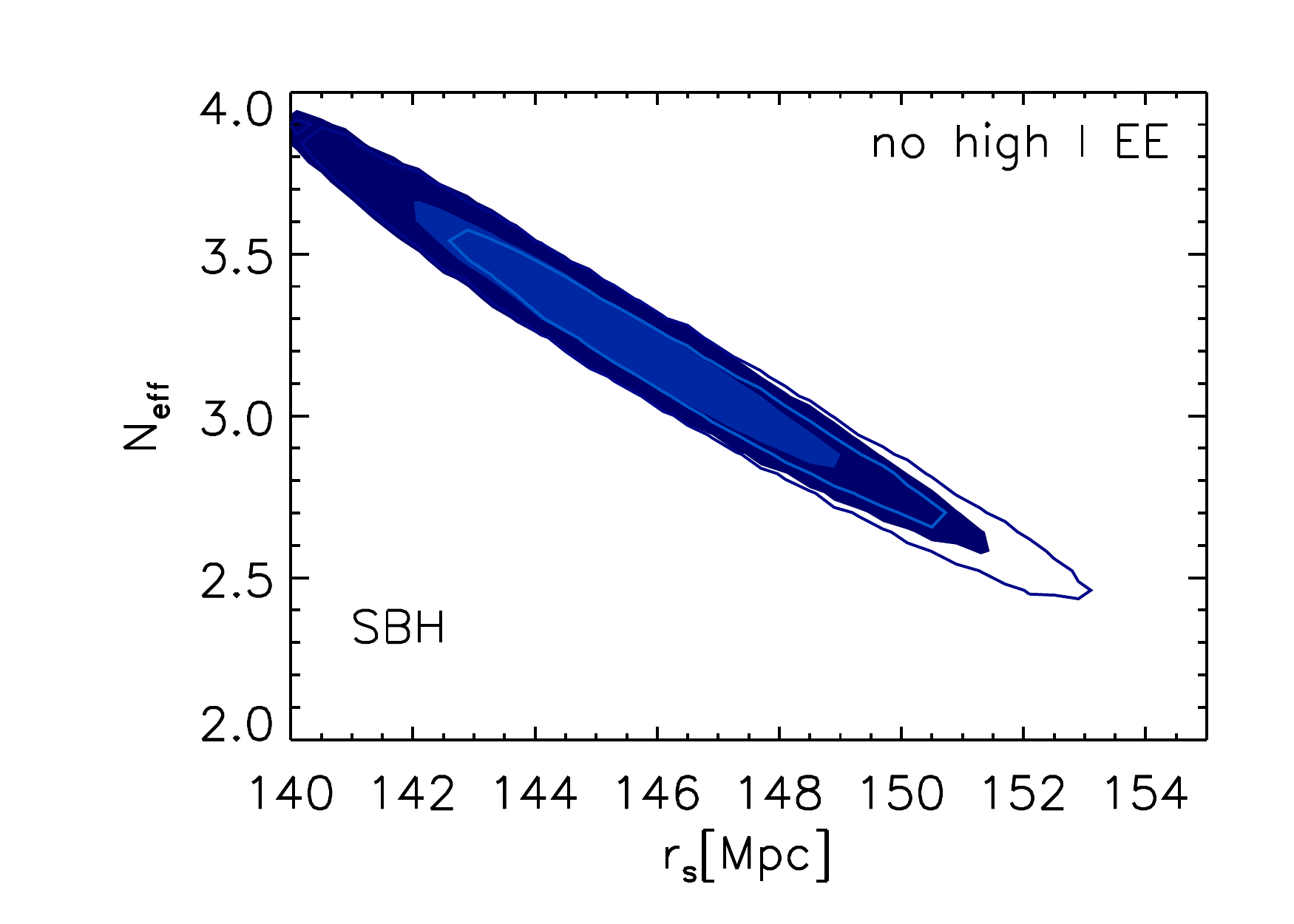}
\includegraphics[width=0.45\textwidth]{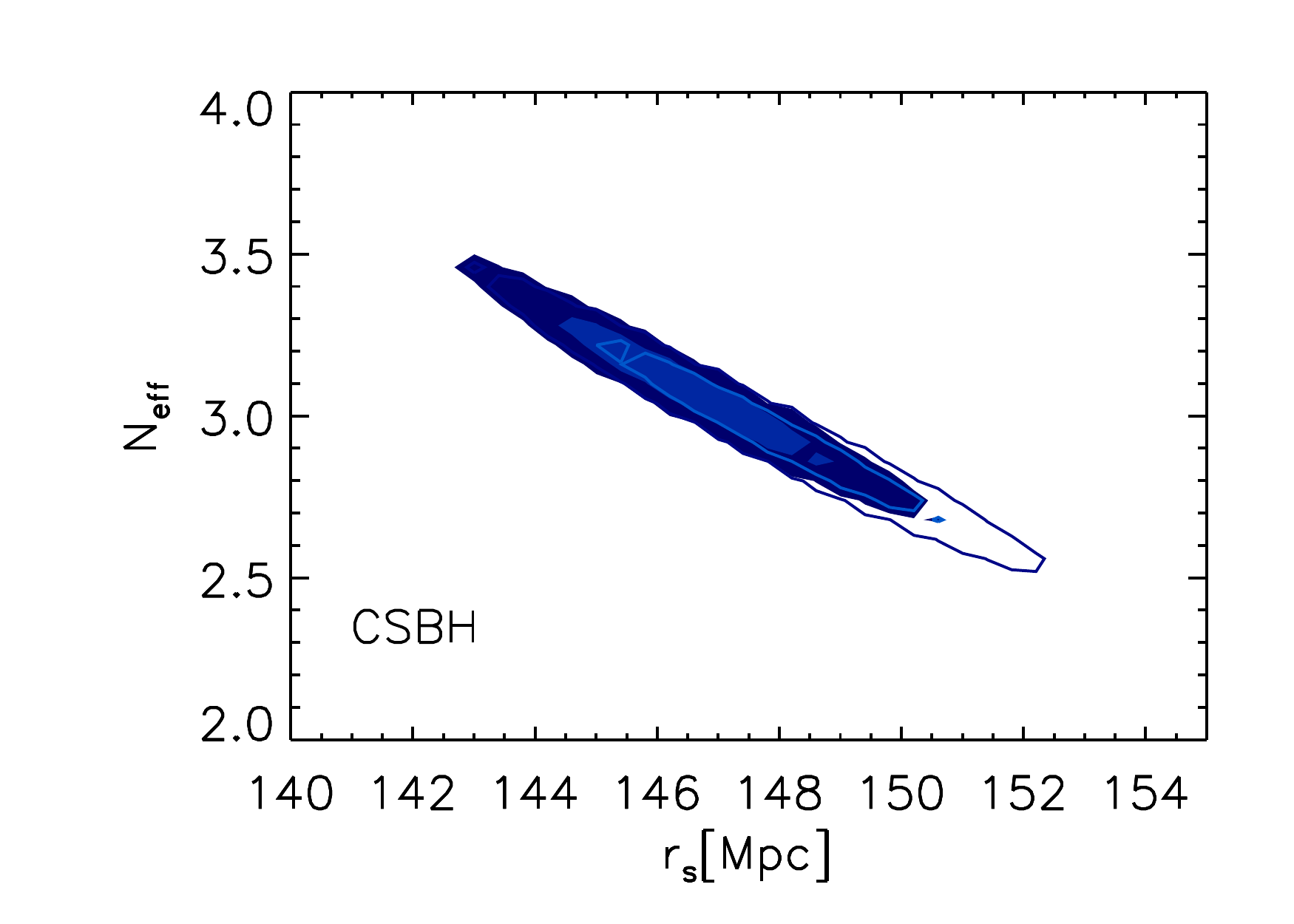}
\includegraphics[width=0.45 \textwidth]{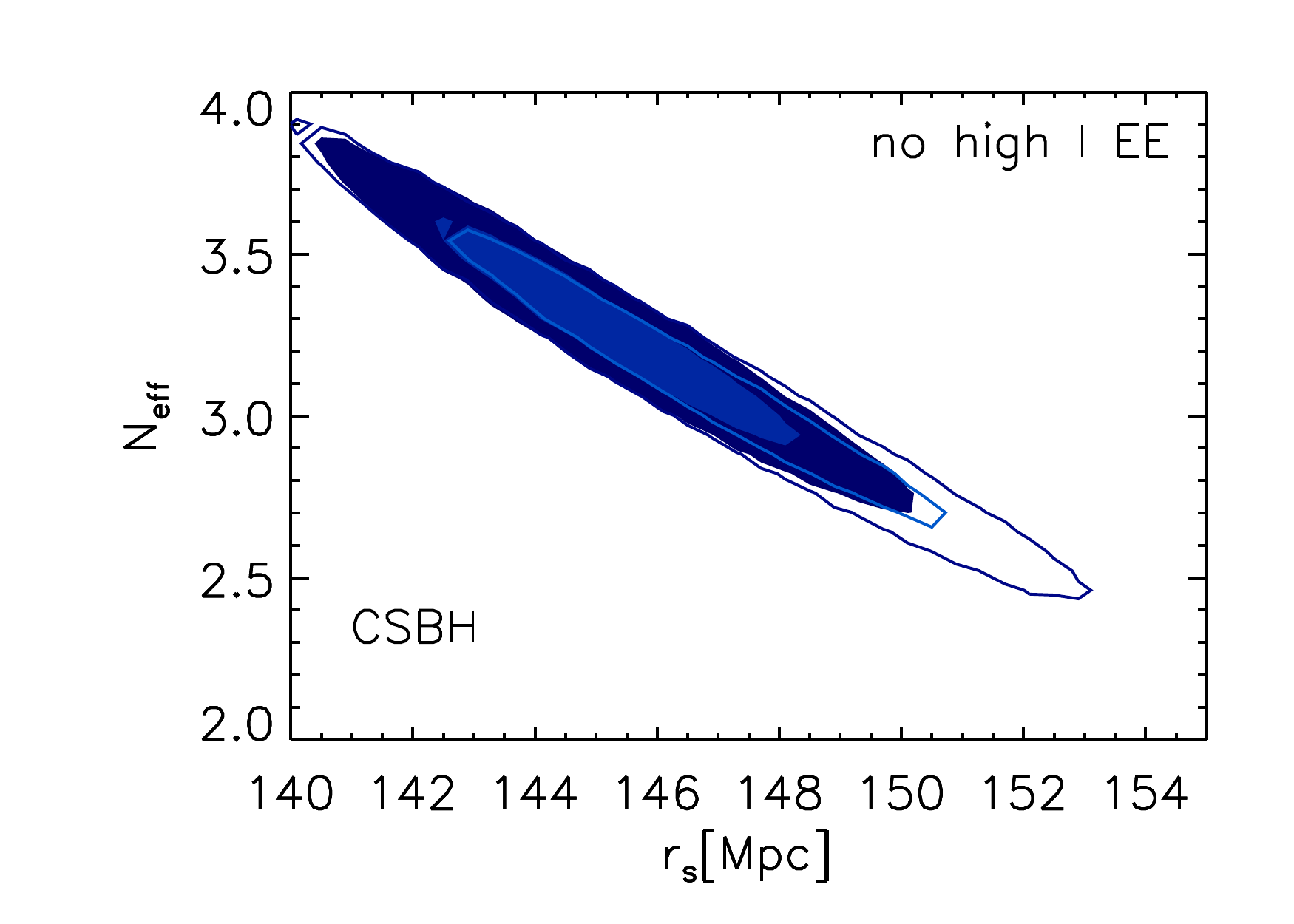}

\caption{Effect of combining the low-redhisft standard ruler measurement (interpreted as the sound horizon at radiation drag) with CMB Planck observations. The transparent contours show the joint $r_{\rm s}$ vs $N_{\rm eff}$ 68\% and 95\% marginalised  confidence regions obtained from the posterior sample provided by  the Planck CMB mission.   On the left, all temperature and polarisation data are used, on the right, high $\ell$ polarisation data are not included.   The filled contours result from importance sampling this with our CSB measurement (top row), SBH (middle row) and CSBH (bottom row).}
\label{fig:neff}
\end{figure*}

\subsection{Robustness to prior assumptions}
\label{sec:robustness}
To assess the dependence on the prior assumptions, we compare our findings with the results and the approach  of \cite{Bernal_H0}(BVR). 
In that work, a similar reconstruction of the late time expansion history is performed in the context of the study of the tension between   the (direct) local  $H_0$ determination and its CMB-inferred  value within the $\Lambda$CDM model. However, they use a different parametrization and sampling method:  $H(z)$ and $r_{\rm s}$ are the  free parameters  and $H(z)$ values are interpolated using natural cubic splines, instead of  $r_{\rm s}^h$,  $\invh(z)$ and linear interpolation as done here. They also use an Affine Invariant sampler (implemented in the public code \texttt{emcee} \citep{emcee}) instead of Metropolis Hastings.  BVR does not include cosmic clocks, so we concentrate on SBH data combination for this test.
The number of nodes is the same ($N=5$), although their location is different. We isolate each of the methodological differences to study their effect in the final results.

As supernov\ae\  data impose very strong constraints on the shape of $H(z)$, the resulting expansion history does not depend on the interpolation method, even taking into account that the splines allow much more freedom than the linear interpolation. Also, the location of the knots does not have any significant effect in the final fit of the reconstruction. It does however have a mild effect on the curvature, which is the parameter most weakly constrained.

In Fig.~\ref{fig:omk_prior} we show the posterior distribution of $\Omega_k$  (left) and  the  joint distribution in the  $\Omega_k$-$r_{\rm s}^h$ plane (right) for the different cases compared in this section. The distributions are marginalised over all other parameters. We refer as `Affine Invariant' to the case when the only change with respect to this work is the MCMC sampler. The figure also quantifies the effect of a different choice of redshift sampling (nodes).
Unlike in our parametrization, using $r_{\rm s}$ and $H(z)$ as free parameters makes  the distribution of  $\Omega_k$ Gaussian,  but centered  around higher  values and with larger error bars.

As $r_{\rm s}$ and $\Omega_k$ are anticorrelated (and $\Omega_k$ and $H_0$ are independent),  differences in the posterior of $\Omega_k$ result in different determinations of the low-redshift standard ruler. The values of $r_{\rm s}$ and $r_{\rm s}^h$ obtained in this work (for the non-flat case) are $\sim 1\sigma$ higher than in BVR. 
Once flatness is imposed, the discrepancies between the two sampling algorithms and prior choices disappear. 

It is important to point out that  the dependence of the posterior on the prior choice and the MCMC sampling method  only appears when  the parameters are  weakly constrained. This is the case when using only BAO, supernov\ae\  
 and $H_0$ (SBH) and not imposing flatness. 
Both  cosmic clocks and supernov\ae\ data are needed to obtain a Gaussian posterior  distribution for the curvature: in these cases (CSBH and CSB), the  dependence on the prior assumptions and the sampler becomes unimportant.  The   dependence on  prior  is negligible  also for  the SBH dataset combination when flatness is imposed.
 
 \section{Discussion and conclusions}

This model-independent determination of the low-redshift standard ruler can be   interpreted as the sound horizon at the baryon drag  and thus compared with  (model-dependent) CMB  determinations. 
This comparison can be used to limit the scope of new physics that may  alter the
early expansion rate and sound speed. This is investigated in detail for example in \cite{edepaper}. Here  we only compare our constraints with those obtained by the Planck team
 with the Planck  2015 data release, using publicly available posterior samples \citep{Planckparameterspaper}. The direct measurement of the ruler is in good agreement with the CMB-derived one for all models considered by the Planck team and especially the standard $\Lambda$CDM model. In all cases the CMB-inferred error bars, are, understandably, much smaller, with one notable exception: the model where the effective number of neutrino species is free \citep{Heavens:2014rja}. The effect of combining our measurement with the CMB one is illustrated in Fig.~\ref{fig:neff}.  Transparent contours are the (joint) 68\% and 95\% confidence regions for CMB data alone including  (excluding) high $\ell$ polarisation data on the left (right) panel.  The filled contours result from importance sampling this with our SBH, CSB or CSBH,  measurement, which reduce the errors significantly. When   $H_0^{SH0ES}$ is included the error on  $N_{\rm eff}$ is reduced by suppressing the posterior for low $N_{\rm eff}$ values. A similar trend was  found by \cite{RiessH0_2016}  and by \cite{Bernal_H0}. 
   
Note that even without an estimate of $h$, the combination of BAO and Supernov\ae\ data already constrain the low-redshift standard ruler scale $r_s^h$ at the 2\% level, $r_s^h=101.0\pm 2.3$ Mpc/$h$.

 Looking ahead, improvements  on the low-redshift standard ruler measurement may arise from the next generation of BAO surveys. For example, if in the CSB (or CSBH) combination we substitute the current BAO measurements with forecasted constraints achievable with a survey with the specifications of DESI \citep{DESI}, errors without imposing flatness will reduce as follows. 
 The error on  $r_{\rm s}^h$ will go from  $1.9$\% to $1.3$\% ($1.8$\% to $1.1$\%), the error on  $r_{\rm s}$ from  3.2\% to 2.8\% (2.2\% to 1.9\%) the error on $H_0$ from   3.7\%  to   3.4\% (2.1\% to 2\%) and the error on $\Omega_k$ from $\pm 0.41$ to $\pm 0.28$  ($\pm 0.31$ to $\pm 0.22$).
Given the dramatic improvement in the precision of expansion history constraints provided by the next generation of BAO surveys, these forecasts  indicate that  we are entering   a regime where the error on $r_{\rm s} $ is dominated by that on  the normalisation of the expansion history $h$, and therefore directly or indirectly on $H_0$.  Improvement on the  local $H_0$ determination towards a goal of $\sim 1$\% error budget may be provided by e.g., gravitational lensing time delays \citep{Suyu} and by further improvements of the classic distance ladder approach \citep{RiessH0_2016}. 

\section{acknowledgements}
LV and JLB thank A. Riess for discussions.
 RJ and LV acknowledge support from Mineco grant  AYA2014-58747-P and MDM-2014-0369 of ICCUB (Unidad de Excelencia Maria de Maeztu), and a visiting scientist grant from the Royal Society. JLB is supported by the Spanish MINECO under grant BES-2015-071307, co-funded by the ESF. JLB and AFH acknowledges hospitality of Radcliffe Institute for Advanced Study, Harvard University.  AFH, RJ and LV acknowledge Imperial College for support under the CosmoCLASSIC collaboration.
 
Based on observations obtained with Planck (http://www.esa.int/Planck), an ESA science mission with instruments and contributions directly funded by ESA Member States, NASA, and Canada.

Funding for SDSS-III has been provided by the Alfred P. Sloan Foundation, the Participating Institutions, the National Science Foundation, and the U.S. Department of Energy Office of Science. The SDSS-III web site is http://www.sdss3.org/.

SDSS-III is managed by the Astrophysical Research Consortium for the Participating Institutions of the SDSS-III Collaboration including the University of Arizona, the Brazilian Participation Group, Brookhaven National Laboratory, Carnegie Mellon University, University of Florida, the French Participation Group, the German Participation Group, Harvard University, the Instituto de Astrofisica de Canarias, the Michigan State/Notre Dame/JINA Participation Group, Johns Hopkins University, Lawrence Berkeley National Laboratory, Max Planck Institute for Astrophysics, Max Planck Institute for Extraterrestrial Physics, New Mexico State University, New York University, Ohio State University, Pennsylvania State University, University of Portsmouth, Princeton University, the Spanish Participation Group, University of Tokyo, University of Utah, Vanderbilt University, University of Virginia, University of Washington, and Yale University.


\end{document}